\documentclass[preprint,showkeys, preprintnumbers,amssymb]{revtex4}

\usepackage{graphicx}
\graphicspath{{figs/}}
\begin{document}
\title{Self-repairing in single-walled carbon nanotubes by heat treatment}
\author{Jin-Wu~Jiang}
\thanks{Electronic mail: phyjj@nus.edu.sg}
	\affiliation{Department of Physics and Centre for Computational Science and Engineering,
 		     National University of Singapore, Singapore 117542, Republic of Singapore }
\author{Jian-Sheng~Wang}
	\affiliation{Department of Physics and Centre for Computational Science and Engineering,
     		     National University of Singapore, Singapore 117542, Republic of Singapore }

\date{\today}
\begin{abstract}
Structure transformation by heat treatment in single-walled carbon nanotubes (SWCNT) is investigated using molecular dynamics simulation. The critical temperature for the collapse of pure SWCNT is as high as 4655 K due to strong covalent carbon-carbon bonding. Above 2000 K, the cross section of SWCNT changes from circle to ellipse. The self-repairing capability is then investigated and two efficient processes are observed for the SWCNT to repair themselves. (1) In the first mechanism, vacancy defects aggregate to form a bigger hole, and a bottleneck junction is constructed nearby. (2) In the second mechanism, a local curvature is generated around the isolate vacancy to smooth the SWCNT. Benefit from the powerful self-repairing capability, defective SWCNT can seek a stable configuration at high temperatures; thus the critical temperature for collapse is insensitive to the vacancy defect density.
\end{abstract}

\maketitle


\section{introduction}
High temperature evaporation is a good method to remove impurities from carbon nanotubes, typically for those samples prepared by the chemical vapor deposition method with Fe/MgO catalysts. This purifying technique initially found its application in the multi-walled carbon nanotubes (MWCNT), and has been developed to clean the double-walled carbon nanotubes (DWCNT) samples recently. In Ref.~\onlinecite{LiuYP}, Liu {\it et~al.} apply the acid dissolution and high temperature evaporation methods to purify nanotube samples. Their experiment shows that the residual Fe particles covered by a carbon shell can not be removed by the chemical solution, but can be cleaned through physical evaporation. A $99\%$ purified sample can be achieved by evaporation treatment at 1700 $^{o}$C. This experiment also demonstrates that the samples with smaller diameter are thermally more stable than that of larger diameter. In Ref.~\onlinecite{KimYA}, Kim {\it et~al.} employ the heat treatment method to investigate the structure transformation of bundles of single-walled carbon nanotubes (SWCNT) at different temperatures. They find that the DWCNT are more stable than SWCNT and even comparable with the MWCNT. They also observe that outer walls of adjacent DWCNT can coalesce into large diameter tubes. SWCNT prefer to change into DWCNT around 2000 $^{o}$C.\cite{YudasakaM, KimUJ2005, KimUJ2008} This is a typical phenomenon in heat treatment experiments for bundled SWCNT. In Ref.~\onlinecite{KimUJ2005, KimUJ2008}, Kim {\it et~al.} demonstrate that the purified bundled SWCNT evolve along very different pathways that depend on three different factors: initial diameter distribution, concomitant tightness of the bundles, and bundle size. Samples prepared by different methods can result in very different final derivatives after heat treatment. These thermal evolutions are claimed to be related to various reasons, such as diffusion of carbon, C-C bond rearrangement, etc.

Theoretically, the heat treatment of carbon nanotubes can be simulated by molecular dynamics (MD). The tight-binding MD is applied to simulate the coalesce of two ultrathin SWCNT after collision, and a conservation of the chiral index is found.\cite{KawaiT} In Ref.~\onlinecite{LopezMJ}, Lopez {\it et~al.} use the extensive MD to study the evolution of the bundles of SWCNT after heat treatment. They find that the bundles of SWCNT coalesce forming MWCNT through an interesting patching-and-tearing mechanism. 

There are two facts we can learn from above experiments and MD simulations. Firstly, these experiments suggest that the SWCNT are thermally unstable after 2000 $^{o}$C and will evolve into DWCNT or MWCNT. Secondly, the underlying mechanism for the evolution of the SWCNT is far from clear. There are lots of possible reasons for this structure evolution. SWCNT samples in most experiments are in bundles and correlated by van der Waals interaction. This interaction facilitates the evolution from SWCNT into DWCNT or MWCNT. Moreover, the residual catalyst impurities are ineluctable in SWCNT samples prepared by the chemical method. These catalysts may assist the evolution of SWCNT. Now a natural question will arise: what will happen if isolated SWCNT undergo heat treatment? How stable are isolated SWCNT? In a MD simulation work for the SWCNT with both open or closed ends, the breakdown temperature $T_{c}$ is found to be about 3300K, which varies in hundreds of Kelvin in different situations.\cite{LiewKM} While in another work, the tight-binding MD is applied and the obtained critical temperature $T_{c}$ is about 2500 K.\cite{DereliG} The large discrepancy between different existing works requires further investigations. This forms one aspect of the topic in this paper.

The defects may play a significant role in the thermal stability of the SWCNT. Among different defects, the vacancy defect can be experimentally generated\cite{EsquinaziP, QuiquiaJB} and well detected.\cite{KimH, FanYW} In Ref.~\onlinecite{EsquinaziP}, the authors apply the proton irradiation to generate vacancy defects in the graphite. The defect density can be quantitatively controlled by applying different dose of the proton. It is estimated that about 0.1$\%$ carbon atoms are displaced for a proton dose of 0.3 nC/$\mu$m$^{2}$. The defect density can be as high as 35$\%$ for a dose of 250 times higher, where the carbon system is amorphous. Considering the advantage of these experimental techniques, it is interesting to study the thermal stability of the SWCNT with different vacancy defects. It has been shown that the vacancy defect can be applied to build a carbon nanotube semiconductor-metal intra molecular junction.\cite{LeeGD} The reconstruction and alignment of vacancies in SWCNT with different diameters are investigated by first-principles and tight-binding calculations.\cite{LeeAT}

In this paper, the evolution of structure configuration of the SWCNT is investigated by MD simulation. The pure isolated SWCNT are thermally stable up to a temperature $T_{c}$=4655 K. The thermal vibration becomes more and more serious with the increase of temperature, and the cross section changes into ellipse at 2000 K. The structure deformation of SWCNT with different vacancy defects is investigated. It is found that the critical temperature $T_{c}$ is insensitive to the vacancy defects. This is resulting from the self-repairing capability of the SWCNT by heat treatment. The vacancy defect can optimize its local environment by exciting phonon modes localized nearby in low temperature region. At high temperatures, vacancy defects can migrate in the SWCNT. There are two efficient self-repairing processes. Firstly, different defects joint together and form a bigger hole; then a bottleneck junction is constructed near the defect. Secondly, local curvature is introduced to smooth the SWCNT. As a result of strong self-repairing capability, the defective SWCNT can find a stable configuration at high temperatures, so the critical temperature $T_{c}$ is insensitive to the defect density.

\section{simulation details}
We run MD simulation using the ``General Utility Lattice Program" (GULP).\cite{Gale} The GULP is usually used to investigate various lattice dynamics properties of crystal systems. There are plenty of reliable interatomic potentials implemented in this code. This is one of the most important features of GULP. Another distinct feature is the use of the symmetric groups of lattice structures to reduce the simulation time considerably. It is worth noting that the MD simulation part in the GULP has been improved significantly in its latest version GULP3.4. Combined with its powerful potential libraries, GULP becomes a useful platform to do MD simulation for the solid crystals. It is suitable for the simulation of structure transition and melting, as some interatomic potentials implemented can characterize the formation and breaking of bonds.

We use the latest version of GULP3.4 for linux system, freely available from its website. The MD simulation is run with canonical ensemble NVT at constant volume and temperature. The constant temperature is realized by the Nos\'{e}-Hoover algorithm.\cite{Nose, Hoover} A small friction parameter of 0.0005 is adopted in the Nos\'{e}-Hoover to minimize the fluctuation in temperature. The frequency of the highest-frequency phonon mode in the SWCNT is about 1600 cm$^{-1}$, corresponding to a vibration period of about 20 fs. We use a time step of 0.001 ps that gives 200 recording data to describe the trajectory in each vibration period of the highest-frequency phonon mode. Usually the total MD simulation time is 2000 ps, about $2\times 10^{6}$ MD steps. The Newton equations are solved by leapfrog Verlet algorithm. We apply the Tersoff potential.\cite{Tersoff} This potential has a good performance in describing the formation and breaking of bonds. We have also tested the Brenner potential.\cite{Brenner} The results are qualitatively the same as Tersoff potential, while Brenner potential is much slower due to its complexity. The structure configurations in this paper are plotted by the software ``Visual Molecular Dynamics"\cite{HumphreyW} with trajectory data from GULP.

\begin{figure}[htpb]
  \begin{center}
    \scalebox{1.3}[1.3]{\includegraphics[width=7cm]{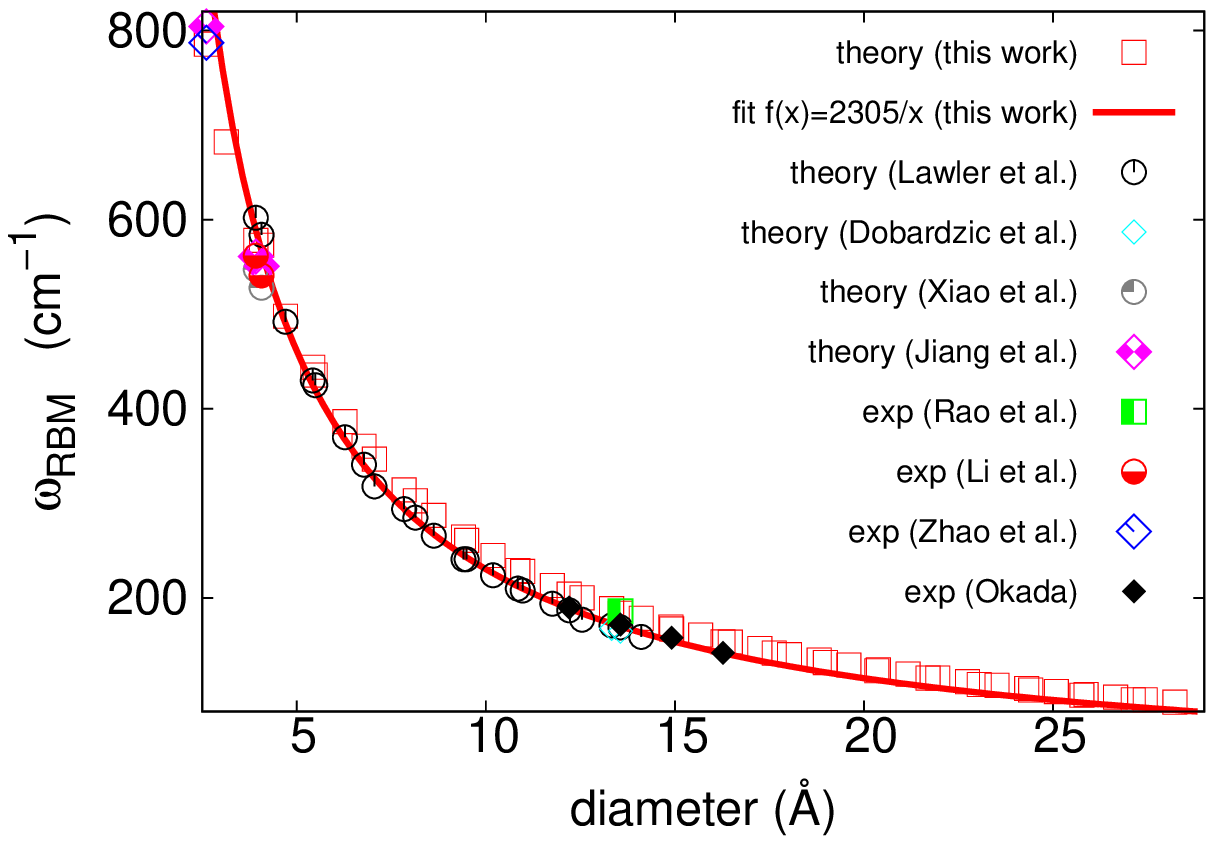}}
  \end{center}
  \caption{(Color online) The frequency of the radial breathing mode in SWCNT. Results from Tersoff potential (this work) are compared with other theoretical or experimental values.}
  \label{fig_rbm}
\end{figure}

Before jumping into the simulation work, it is meaningful for us to check the suitability of the Tersoff potential for interatomic interactions in SWCNT. Tersoff potential has found various applications in the description of covalent bonds. For example, this potential is applied to study the thermal conductance at interface between SWCNT and silicon,\cite{DiaoJ} the structure of SWCNT under hydrostatic pressure,\cite{ImtaniAN} and the H$_{2}$ adsorption effect on the SWCNT.\cite{OkatiA} As an additional check, we use the Tersoff potential to calculate the radial breathing mode (RBM) in the SWCNT. On the one hand, the RBM reflects the carbon-carbon bond strength in the SWCNT, thus is a good candidate for the check of the potential. On the other hand, the RBM in SWCNT has been extensively studied, so there are lots of existing results to compare with. RBM is widely used as a fundamental tool to clarify the diameter of the tubes experimentally, because it is a Raman active mode and its frequency is inversely proportional to the tube diameter.\cite{RaoAM, LiZM, ZhaoX, OkadaS} This mode was theoretically investigated by {\it ab initio} calculation,\cite{LawlerHM} and different valence force field models.\cite{Dobardzic, XiaoY, JiangJW} In Fig.~\ref{fig_rbm}, we show the frequency of RBM with diameter from 2.5~{\AA} to 28~{\AA}, including armchair SWCNT (n, n) with $2\le n \le 20$ and zigzag SWCNT (n, 0) with $4 \le n \le 36$. Our results are in good agreement with both existing experimental data and theoretical values in whole diameter range. We need to stress that the Tersoff potential gives good results not only for the wider SWCNT but also for narrow SWCNT with diameter below 4.0~{\AA}. This shows the advantage of the Tersoff potential for SWCNT, because it is not an easy job to obtain correct frequency for RBM in narrow SWCNT, even for the {\it ab initio} calculation. Another important fact is that the frequency of RBM from Tersoff potential can be fitted as a function of diameter $\omega_{\rm RBM}=A/d$ where $d$ is the tube diameter. The fitting parameter $A=2305 \pm 24$ cm$^{-1}${\rm \AA} is very close to the first principle calculated value of 2260 cm$^{-1}$\AA.\cite{LawlerHM} All these results confirm the suitability of the Tersoff potential for the description of interatomic interactions in SWCNT.

\section{results and discussion}
\begin{figure}[htpb]
  \begin{center}
    \scalebox{1.2}[1.2]{\includegraphics[width=7cm]{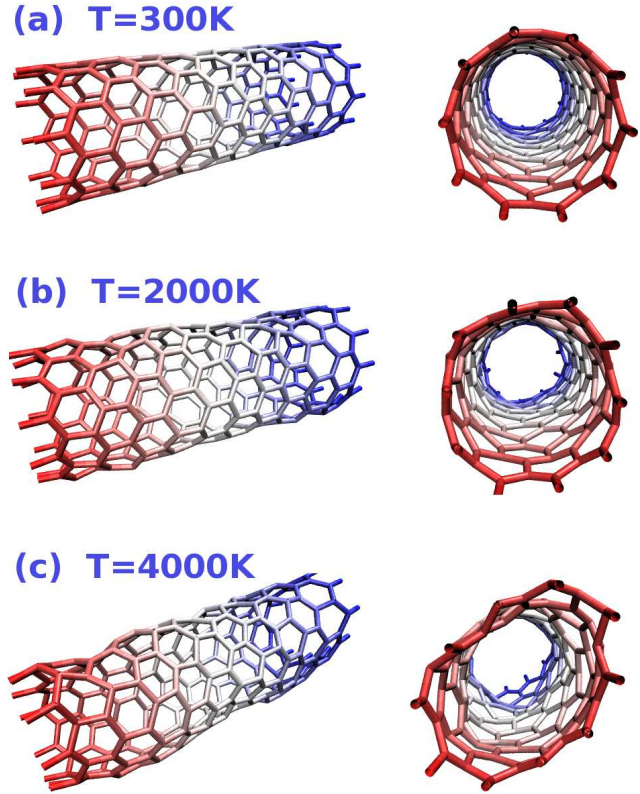}}
  \end{center}
  \caption{(Color online) Structure of a perfect SWCNT(10, 0) at different temperatures. Left/right panel are the side/top view. The colors are with respect to the position of each atom and used to make the figure more clear. The colors in following figures are the same.}
  \label{fig_pure_nodestroy}
\end{figure}

We study the structure evolution of the zigzag SWCNT (10, 0) with the length of 26~{\AA} at different temperatures. There are similar phenomena observed in armchair SWCNT (10, 10). Periodic boundary condition is applied in the axial direction. Large periods are applied in the two lateral directions, so that the tube is isolated equivalently in these two directions. Fig.~\ref{fig_pure_nodestroy} shows that the thermal vibration is almost invisible at 300 K, because the C-C covalent bonds in the SWCNT are very strong, typically in the order of 2.0 eV. The thermal vibration becomes obvious at 2000 K, where the cross section is not a perfect circle anymore. One interesting result is that the tube is far from being destroyed at 2000 K. In some experiments on the bundles of SWCNT,\cite{YudasakaM, KimUJ2005, KimUJ2008} it was found that SWCNT will always turn into MWCNT above 2000 K. It is possible to attribute this experimental phenomenon to van der Waals interaction between different SWCNT, and much more complicated environment in the experiments. Our simulation result indicates that an isolated SWCNT in a clean environment is still stable at 2000 K. However, strong thermal vibration and deformation can be observed in this temperature range.

\begin{figure}[htpb]
  \begin{center}
    \scalebox{1.2}[1.2]{\includegraphics[width=7cm]{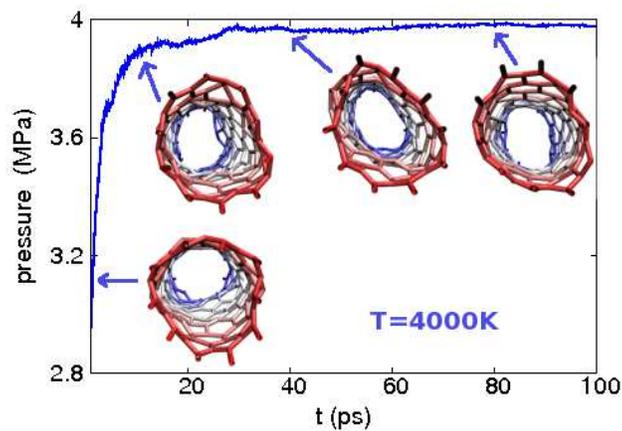}}
  \end{center}
  \caption{(Color online) Time evolution of pressure at 4000 K. The structure transition for SWCNT(10, 0) is illustrated as insets through some configurations of the SWCNT.}
  \label{fig_P_T_4000}
\end{figure}

We learn from Fig.~\ref{fig_pure_nodestroy} that the thermal vibration becomes more and more serious with the increase of temperature. All carbon atoms are no longer in the same cylindrical surface at high temperatures. We always observe that the middle part of the tube and the two ends of the tube are deformed in perpendicular directions. Carbon atoms are still not evaporated at 4000 K, although they are in an extremely strong thermal vibration. There is no 5- or 7-membered rings for SWCNT observed here. Fig.~\ref{fig_P_T_4000} is the time evolution of internal pressure at 4000 K. The pressure is calculated by:\cite{MelchionnaS, MartynaGJ}
\begin{eqnarray}
P=\frac{1}{VD}\left[\sum_{i=1}^{N}\frac{p_{i}^{2}}{m_{i}} + \sum_{i=1}^{N}\vec{r}_{i}\cdot\vec{F}_{i} - (VD)\frac{\partial \phi (\vec{r},V)}{\partial V}  \right]
\end{eqnarray}
where $D$ is the degree of freedom and $N$ is the atom number. $\phi$ is the interatomic interaction which is Tersoff potential here. This is the internal pressure and is used to balance with the externally applied pressure in the thermal steady state. This curve can monitor structure evolution of a tube. The pressure will have a rapid change if the tube changes from one configuration into another. Insets are some typical configurations at different simulation stages.

\begin{figure}[htpb]
  \begin{center}
    \scalebox{1.2}[1.2]{\includegraphics[width=7cm]{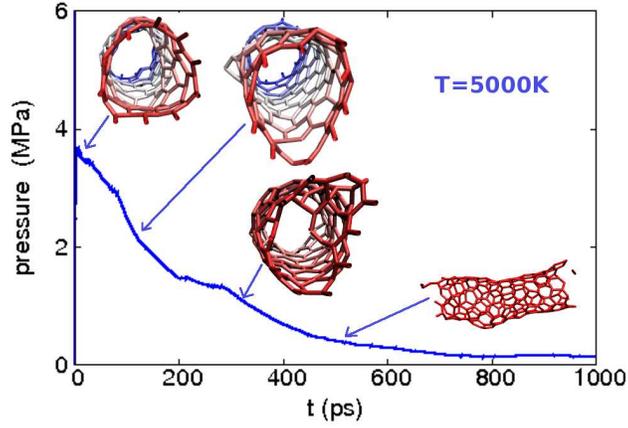}}
  \end{center}
  \caption{(Color online) Time evolution of pressure at 5000 K. The structure of SWCNT(10, 0) is finally destroyed. Insets are some typical configurations of the SWCNT during collapse.}
  \label{fig_pure_destroy}
\end{figure}

SWCNT will be destroyed into pieces at temperatures above a critical value $T_{c}=4655$ K. Fig.~\ref{fig_pure_destroy} illustrates the collapse of SWCNT step by step. Usually the collapse of a tube starts from the damage of a particular region, where some atoms are evaporated. Then some atoms in other areas will move into this region and be evaporated again. The tube is finally destroyed totally. The critical temperature $T_{c}=4655$ K is higher than two previous theoretical works,\cite{LiewKM, DereliG} where the value is about 3300 K and 2500 K. One possible reason for this difference is the different interaction potentials which affect the value of $T_{c}$ directly. In these two works, the total energy is either described by Brenner potential,\cite{LiewKM} or calculated in the tight-binding method.\cite{DereliG} Another possible reason for this difference is the simulation time step. A large time step will accumulate numerical errors in the solution of the Newton equations in MD simulation. Some atoms will leave the tube solely due to the numerical error, leading to the collapse of the SWCNT. For example, we can artificially increase the time step to be 0.003 ps. It results in a value of 3600 K for the critical temperature. Obviously, this value is not the actual critical temperature for the tube. The boundary conditions in the axial direction is also very important. There are bending modes in the tube if both ends are fixed. These bending modes will facilitate the breakdown of the tube from its middle region.\cite{LiewKM}

For a longer tube with length 52~{\AA}, we find that the critical temperature is almost the same as the shorter one. This is because we apply the periodic boundary condition in the axial direction. The periodic length is large enough, so there is no much difference with the increase of length in the axial direction. For a wider tube (20, 0) with the same length 26~{\AA}, it is interesting that the $T_{c}$ is significantly lower, about 4000 K. This result is qualitatively consistent with the experiment,\cite{LiuYP} where tubes in smaller diameter are thermally more stable, although the curvature energy in smaller tubes is larger.\cite{GulserenO} The curvature energy is the internal strain energy of a SWCNT, by viewing SWCNT as rolling up a graphene sheet into cylindrical structure.\cite{WhiteCT}

Up to now, we have concentrated on pure SWCNT without any defects. From now on, we will switch to the discussion of the heat treatment of SWCNT with vacancy defects. A vacancy defect is introduced by arbitrarily removing one carbon atom from the pure SWCNT. Fig.~\ref{fig_defect1} shows similar results as pure tubes at low temperature (300 K), where thermal vibration is invisible. At 2000 K, atoms around the defect will optimize their structure to lower the local energy introduced by the defect. These atoms have obvious larger vibrational amplitudes than the other atoms, which results from the localized phonon modes around this defect. If these localized phonon modes are excited, atoms around the defect have strong vibration while the other atoms almost do not move. At 4000 K, we can see that the defect starts to move in the tube. Fig.~\ref{fig_defect_T_4000} displays how the defect shifts in a SWCNT. Finally, the defect will settle down in some region, and optimize the defect to an energy stable state. With the further increase of temperature, some atoms will be evaporated in the defective SWCNT but tubes are still stable. The critical temperature for SWCNT with only one defect is about 4635 K, which is almost the same as the corresponding pure SWCNT.

To see the self-repairing process more clearly, we need to introduce more defects in the SWCNT. We have tried two, four, eight, sixteen, and thirty-two defects. There are totally 240 atoms in this studied SWCNT, and the corresponding defect density for two defects is 0.83\%. It is interesting that the critical temperature $T_{c}$ for these defective tubes is roughly the same as pure ones, around 4600 K. The fluctuation is just some tens of Kelvin. At first glance, the defects should result in much lower critical temperature. However, this result can be considered in another aspect. A stable value of $T_{c}$ implies the strong self-repairing capability of the SWCNT. In other words, the SWCNT can repair the defects by itself and turns into a stable configuration. Fig.~\ref{fig_defect_repair} shows how this self-repairing is accompanied in the SWCNT in case of eight defects. There are two efficient repairing methods. In the first process (denoted by double blue arrow), two defects move closer and closer, and finally get together to form a bigger hole. Then a bottleneck junction is constructed near this hole. It looks like a junction structure between two tubes with different diameters. In another repairing method (denoted by single red arrow), two defects get closer and the neighborhood around these two defects is smoothed by introducing a local curvature. Usually, the bottleneck junction favors to occur around a bigger defect; while the smoothing mechanism always happens between many smaller defects. These two mechanisms are not independent of each other. They can happen continuously. For example, a bottleneck junction is formed and immediately further smoothed. In case of large defect density, these two self-repairing mechanisms will cooperate with each other. We find that some defects will always stay along and smoothed by the heat treatment. While some defects will get together and form a bigger hole, which will form a bottleneck junction nearby. We want to further remark that the smoothing mechanism can work for self-repairing, and this repair will introduce a curvature surface around the defect as a byproduct. For a small hole defect, a slight curvature is enough to smooth the area around the defect. However, for a big hole defect, large curvature is in need to achieve a more stable structure by the smoothing mechanism. The curvature surface introduced by the smoothing mechanism is so large that it eventually turns into a bottleneck junction.

\section{conclusion}
To conclude, MD simulation is performed to investigate the structure deformation of SWCNT by heat treatment. It is found that the critical temperature for the collapse of the pure isolated SWCNT is about 4655 K, which results from the strong covalent C-C bonding. With the increase of temperature, the cross section of SWCNT changes into ellipse at 2000 K. The self-repairing processes in SWCNT with vacancy defects are studied in detail. There are two efficient mechanisms for the SWCNT to do the self-repairing work. The first one is to construct a bottleneck junction after two defects joining together. Another process is to introduce a local curvature to smooth the surface around the defect. Due to strong capability of self-repairing, SWCNT can find a stable configuration even at very high temperature. As a result, the critical temperature for the collapse of SWCNT is insensitive to the defect density. We expect that this self-repairing phenomenon can be observed in the experiment, since the generation of vacancy defects can be quantitatively controlled,\cite{EsquinaziP, QuiquiaJB} and high vacuum chamber technique is also at a high level.

\section*{Acknowledgements}
The work is supported by a Faculty Research Grant of R-144-000-257-112 of National University of Singapore.

\begin{widetext}
\begin{figure}[htpb]
  \begin{center}
    \scalebox{1.2}[1.2]{\includegraphics[width=7cm]{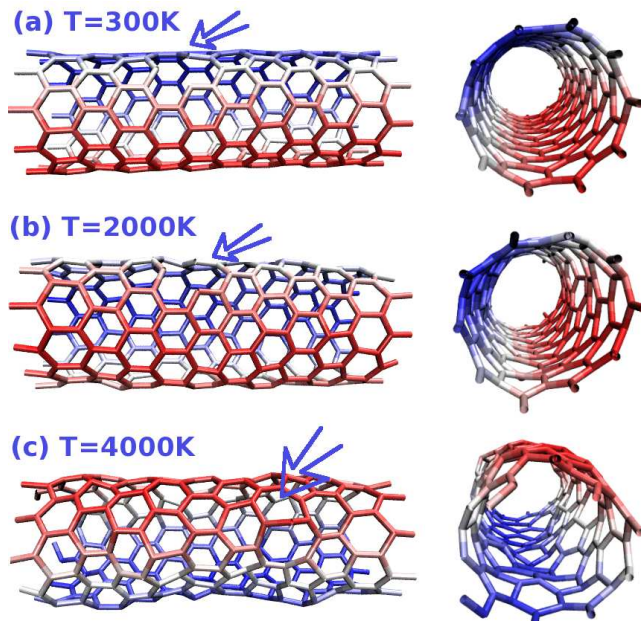}}
  \end{center}
  \caption{(Color online) Structure of SWCNT(10, 0) with one defect at different temperatures. Left/right panel are the side/top view.}
  \label{fig_defect1}
\end{figure}

\begin{figure}
  \includegraphics[width=14cm]{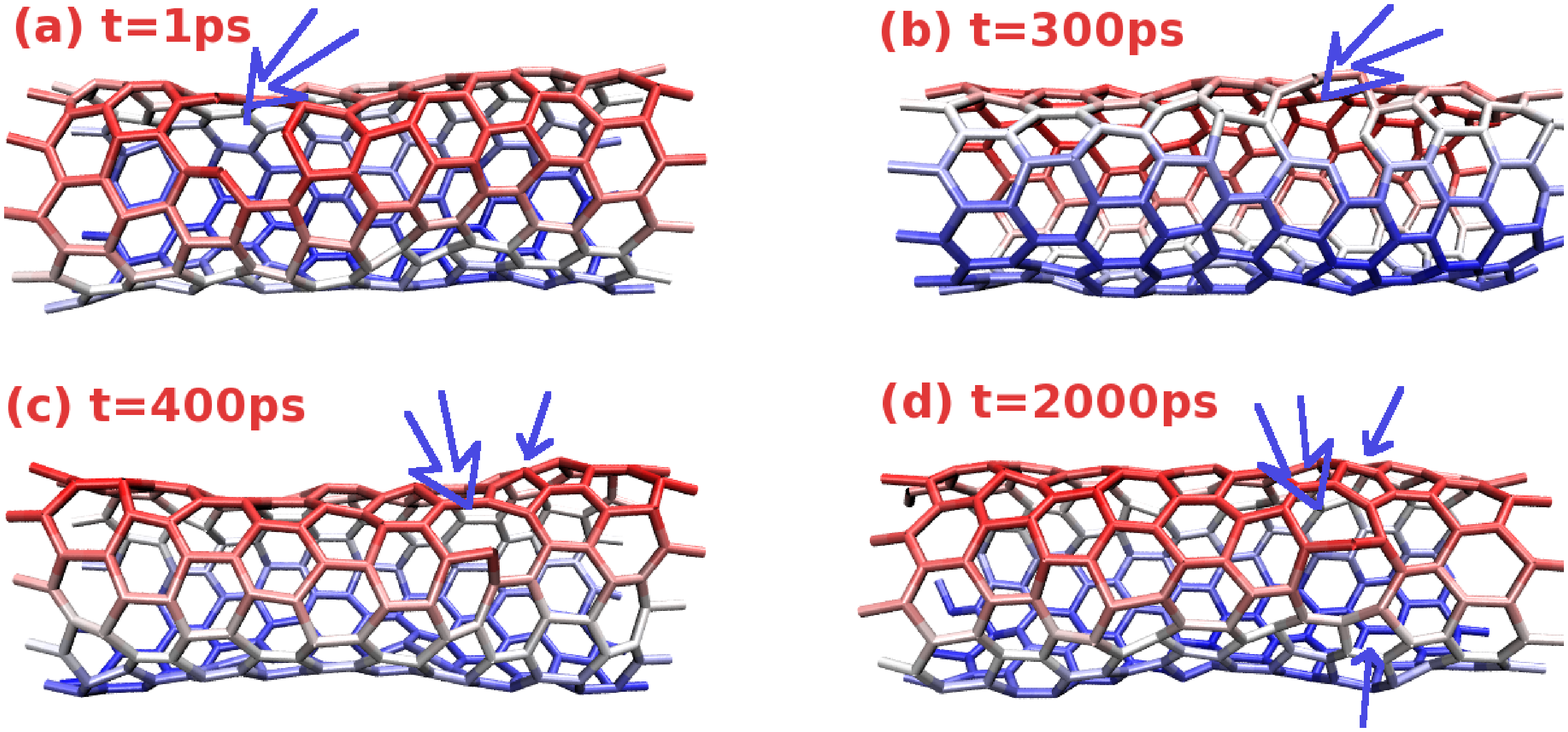}
  \caption{(Color online) Migration of the defect (denoted by arrows) in the SWCNT at 4000 K.}
  \label{fig_defect_T_4000}
\end{figure}
\begin{figure}
  \includegraphics[width=14cm]{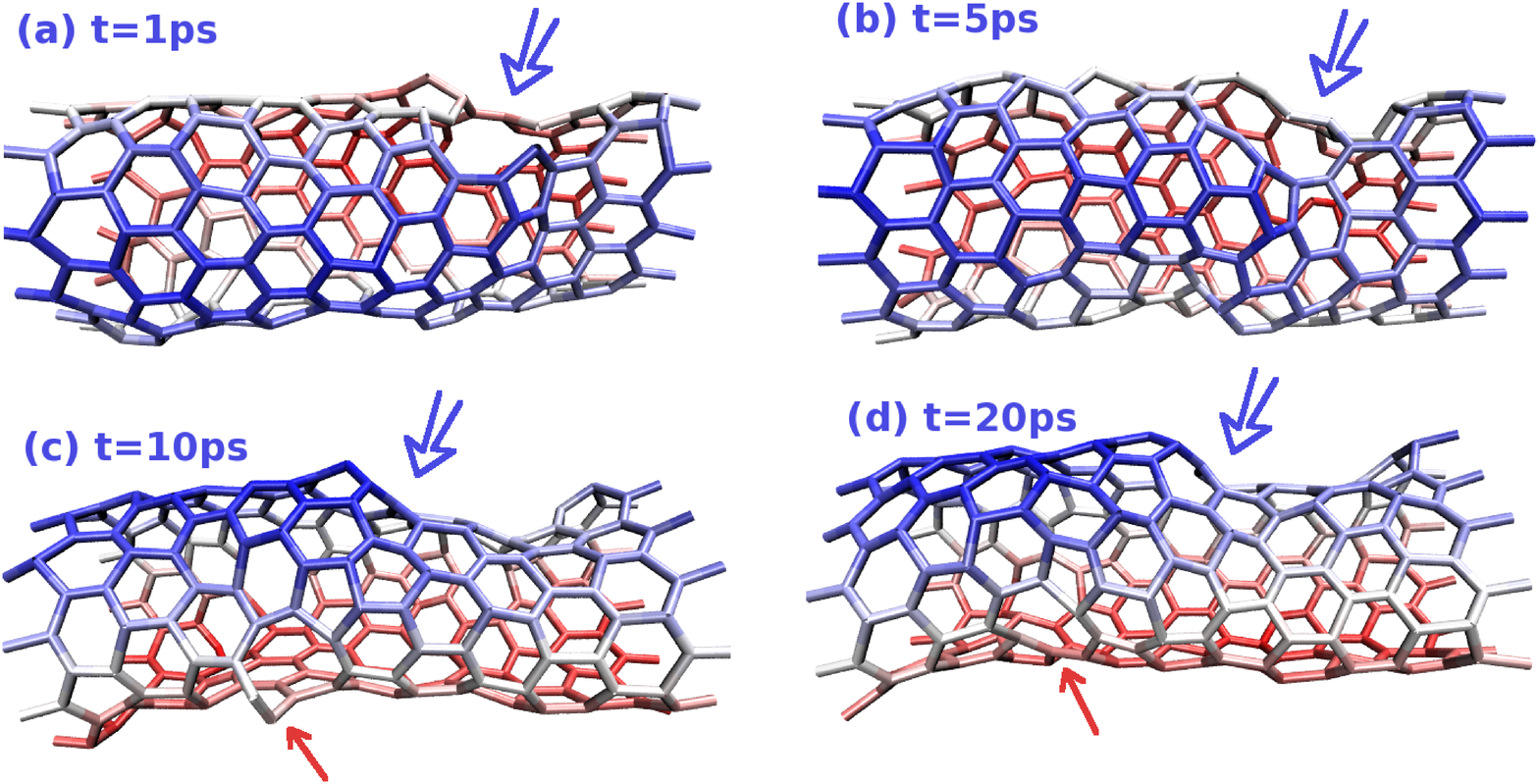}
  \caption{(Color online) Self-repairing in the SWCNT with eight defects at 4000 K. Two repairing mechanisms: (1) bottleneck formation denoted by double blue arrow; (2) smoothing mechanism denoted by single red arrow.}
  \label{fig_defect_repair}
\end{figure}

\end{widetext}


\begin{thebibliography}{}
\bibitem{LiuYP} Y. P. Liu, Y. Wang, Y. Liu, W. J. Li, W. P. Zhou, and F. Wei, Nanotechnology \textbf{18}, 175704 (2007).

\bibitem{KimYA} Y. A. Kim, H. Muramatsu, T. Hayashi, M. Endo, M. Terrones, and M.S. Dresselhaus, Chem. Phys. Lett. \textbf{398}, 87 (2004).

\bibitem{YudasakaM} M. Yudasaka, T. Ichihashi, D. Kasuya, H. Kataura, and S. Iijima, Carbon \textbf{41} 1273 (2003).

\bibitem{KimUJ2005} U. J. Kim, H. R. Gutierrez, J. P. Kim, and P. C. Eklund, J. Phys. Chem. B \textbf{109}, 23358 (2005).

\bibitem{KimUJ2008} U. J. Kim, H. R. Gutierrez, A. K. Guptaa, and P. C. Eklund, Carbon \textbf{46}, 729 (2008).

\bibitem{KawaiT} T. Kawai, Y. Miyamoto, O. Sugino, and Y. Koga, Phys. Rev. Lett. \textbf{89}, 085901 (2002).

\bibitem{LopezMJ} M. J. Lopez, A. Rubio, J. A. Alonso, S. Lefrant, K. Metenier, and S. Bonnamy,  Phys. Rev. Lett. \textbf{89}, 255501 (2002).

\bibitem{LiewKM} K. M. Liew, C. H. Wong, X. Q. He, and M. J. Tan, Phys. Rev. B \textbf{71}, 075424 (2005).

\bibitem{DereliG} G. Dereli, B. S\"{u}ng\"{u}, and C. \"{O}zdogan, Nanotechnology \textbf{18}, 245704 (2007).

\bibitem{EsquinaziP} P. Esquinazi, D. Spemann, R. Hohne, A. Setzer, K. H. Han, and T. Butz, Phys. Rev. Lett. \textbf{91}, 227201 (2003).

\bibitem{LeeGD} G. D. Lee, C. Z. Wang, J. Yu, E. Yoon, N. M. Hwang, and K. M. Ho, Phys. Rev. B \textbf{76}, 165413 (2007).

\bibitem{LeeAT} A. T. Lee, Y. J. Kang, K. J. Chang, and I. H. Lee, Phys. Rev. B \textbf{79}, 174105 (2009).

\bibitem{QuiquiaJB} J. B. Quiquia, P. Esquinazi, M. Rothermel, D. Spemann, T. Butz, and N. Garcia, Phys. Rev. B \textbf{76}, 161403(R) (2007).

\bibitem{KimH} H. Kim, J. Lee, S. J. Kahng, Y. W. Son, S. B. Lee,  C. K. Lee, J. Ihm, and Y. Kuk, Phys. Rev. Lett. \textbf{90}, 216107 (2003).

\bibitem{FanYW} Y. W. Fan, B. R. Goldsmith, and P. G. Collins, Nature Mater. \textbf{4}, 906 (2005).

\bibitem{Gale} J. D. Gale, JCS Faraday Trans., \textbf{93}, 629 (1997).

\bibitem{Nose} S. Nos\'{e}, J. Chem. Phys. \textbf{81}, 511 (1984).

\bibitem{Hoover} W. G. Hoover, Phys. Rev. A, \textbf{31}, 1695 (1985).

\bibitem{Tersoff} J. Tersoff, Phys. Rev. B \textbf{38}, 9902 (1988).

\bibitem{Brenner} D. W. Brenner, O. A. Shenderova, J. A. Harrison, S. J. Stuart, B. Ni, and S. B. Sinnott, J. Phys.:Condens. Matter \textbf{14}, 783 (2002).

\bibitem{HumphreyW} W. Humphrey, A. Dalke, and K. Schulten, J. Molec. Graphics, \textbf{14}, 33 (1996).

\bibitem{DiaoJ} J. Diao, D. Srivastava D, and M. Menon, J. Chem. Phys. \textbf{128}, 164708 (2008).

\bibitem{ImtaniAN} A. N. Imtani and V. K. Jindal, Comput. Mater. Sci. \textbf{46}, 297 (2009).

\bibitem{OkatiA} A. Okati, A. Zolfaghari, F. S. Hashemi, N. Anousheh, and H. Z. Jooya, Fuller. Nanotub. Car. N. \textbf{17}, 324 (2009).

\bibitem{RaoAM} A. M. Rao, E. Richter, S. Bandow, B. Chase, P. C. Eklund, K. A. Williams, S. Fang, K. R. Subbaswamy, M. Menon, A. Thess, R. E. Smalley, G. Dresselhaus, and M. S. Dresselhaus, Science \textbf{275}, 187 (1997).

\bibitem{LiZM} Z. M. Li, Z. K. Tang, G. G. Siu, and I. Bozovic, Appl. Phys. Lett. \textbf{84}, 4101 (2004).

\bibitem{ZhaoX} X. Zhao, Y. Liu, S. Inoue, T. Suzuki, R. Jones, and Y Ando, Phys. Rev. Lett. \textbf{92}, 125502 (2004).

\bibitem{OkadaS} S. Okada, Chem. Phys. Lett. \textbf{438}, 59 (2007).

\bibitem{LawlerHM} H. M. Lawler, D. Areshkin, J. W. Mintmire, and C. T. White, Phys. Rev. B \textbf{72}, 233403 (2005).

\bibitem{Dobardzic} E. Dobard\u{z}i\'{c}, I. Milo\u{s}evi\'{c}, B. Nikoli\'{c}, T.Vukovi\'{c}, and M. Damnjanovi\'{c}, Phys. Rev. B \textbf{68}, 045408 (2003).

\bibitem{XiaoY} Y. Xiao, Z. M. Li, X. H. Yan, Y. Zhang, Y. L. Mao, and Y. R. Yang, Phys. Rev. B \textbf{71}, 233405 (2005).

\bibitem{JiangJW} J. W. Jiang, H. Tang, B. S. Wang, and Z. B. Su, J. Phys.: Condens. Matter \textbf{20}, 045228 (2008).

\bibitem{MelchionnaS} S. Melchionna, G. Ciccotti, B. L. Holian, Mol. Phys. \textbf{78}, 533, (1993).

\bibitem{MartynaGJ} G. J. Martyna, D. J. Tobias, and M. L. Klein, J. Chem. Phys. \textbf{101}, 4177 (1994).

\bibitem{GulserenO} O. Gulseren, T. Yildirim, and S. Ciraci, Phys. Rev. B \textbf{65}, 153405 (2002).

\bibitem{WhiteCT} C. T. White, D. H. Robertson, and J. W. Mintmire, Phys. Rev. B \textbf{47}, 5485 (1993).

\end{thebibliography}
\end{document}